\documentclass[a4paper,11pt]{article}
\usepackage{pos}

\usepackage{subcaption}

\makeatletter
\providecommand*{\diff}%
  {\@ifnextchar^{\DIfF}{\DIfF^{}}}
\def\DIfF^#1{%
  \mathop{\mathrm{\mathstrut d}}%
    \nolimits^{#1}\gobblespace}
\def\gobblespace{%
    \futurelet\diffarg\opspace}
\def\opspace{%
    \let\DiffSpace\!%
    \ifx\diffarg(%
      \let\DiffSpace\relax
     \else
      \ifx\diffarg[%
	\let\DiffSpace\relax
      \else
	\ifx\diffarg\{%
	  \let\DiffSpace\relax
	\fi\fi\fi\DiffSpace}
\makeatother
	
\title{Radiative corrections for the decay $\Sigma^0\to\Lambda e^+e^-$}

\makeatletter
\def\@fnsymbol#1{\ensuremath{\ifcase#1\or \dagger\or \ddagger\or \mathsection\or \mathparagraph\or \|\or **\or \dagger\dagger \or \ddagger\ddagger \else\@ctrerr\fi}}
\makeatother

\author*[\ddagger]{Tom\'a\v{s} Husek}

\affiliation{Institute of Particle and Nuclear Physics, Faculty of Mathematics and Physics, Charles University,\\
V Hole\v{s}ovi\v{c}k\'ach 2, 18000 Praha 8, Czech Republic%
\footnotemark[0]\footnotetext[0]{%
\hspace{-2.2mm}%
\setlength{\tabcolsep}{2pt}%
\begin{tabular}{rl}
$^\ddagger$Present address: & Department of Astronomy and Theoretical Physics, Lund University,\\
&S\"olvegatan 14A, SE 223-62 Lund, Sweden\end{tabular}}%
}

\emailAdd{tomas.husek@thep.lu.se}

\abstract{
Electromagnetic form factors serve to explore the intrinsic structure of nucleons and their strangeness partners.
For the decay $\Sigma^0\to\Lambda e^+e^-$, due to limited phase space, the effects caused by the Sigma-to-Lambda transition form factors compete with the QED radiative corrections.
These corrections are addressed in the present work, evaluated beyond the soft-photon approximation, i.e., over the whole range of the Dalitz plot and with no restrictions on the energy of the radiative photon.
}

\FullConference{%
  40th International Conference on High Energy physics - ICHEP2020\\
  July 28 - August 6, 2020\\
  Prague, Czech Republic (virtual meeting)
}

\begin{document}
\maketitle

\section{Introduction}

One of the major challenges in contemporary particle physics is to understand confinement: How are matter building blocks (quarks and gluons) distributed within composite objects we call hadrons.
If we take, as an example, nucleons, we can study their properties by scattering electrons off them, measure electromagnetic form factors and extract further relevant low-energy hadronic quantities like electric and magnetic radii.
Another way how to study composite objects would be modifying one of their components and observe the resulting objects and their properties.
For nucleons, if one replaces one or both down quarks with strange quark(s), studying the resulting hyperons might provide us with complementary information to what we already know about nucleons.

Performing experiments involving hyperons while using a corresponding setting turns out to be a rather difficult task (at least, as compared to the nucleon case) due to their instability.
In any case, both transition and direct electromagnetic form factors can be measured on the electron--positron colliders, although only above the hyperon--antihyperon pair production threshold.
On the other hand, in the low-energy region, the instability of hyperons can be turned into an advantage and we might study their Dalitz decays, which allow us to access information about the form factors at vanishing photon virtuality.

One such an example of the Dalitz decay in the ground-state baryon-octet sector is the process $\Sigma^0\to\Lambda e^+e^-$, which allows us to study the electromagnetic form factor of the $\Sigma^0\to\Lambda$ transition in a small virtuality window of $\approx$77\,\text{MeV}.
Extracting hadronic quantities like electric and magnetic radii is of a special interest, since it would be compelling to compare these measurement with theoretical predictions.
However, such an endeavour faces several obstacles. Not only are the radii predicted to be rather small parameters~\cite{Kubis:2000aa,Granados:2017cib}, but their effect is further suppressed by a limited phase space.
Therefore, not only high-precision and high-statistics measurement is required to achieve this goal (such opportunities are expected with an advent of future hyperon factories), 
but it becomes clear that the hadronic effects compete in size with electromagnetic radiative corrections.
The calculation of the inclusive NLO QED radiative corrections is the topic of the presented talk.

\section{Leading order}

Before we get into radiative corrections in greater detail, let us first introduce our notation on a simple calculation of the leading-order contribution.
For the $\Sigma^0\Lambda\gamma$ vertex we write
\begin{equation}
\langle0|j^\mu|\Sigma^0\bar\Lambda\rangle
=e\bar v_\Lambda(\vec p_{\Lambda})\,G^\mu(p_{\Sigma^0}+p_{\Lambda})\,u_\Sigma(\vec p_{\Sigma^0})\,,
\end{equation}
with
\begin{equation}
G^\mu(q)
\equiv
\bigg[\gamma^\mu-(M_{\Sigma^0}-M_\Lambda)\frac{q^\mu}{q^2}\bigg]G_1\big(q^2\big)
-\frac{i\sigma^{\mu\nu}q_\nu}{M_{\Sigma^0}+M_\Lambda}G_2\big(q^2\big)\,,
\end{equation}
where we introduced the Dirac and Pauli form factors $G_1$ and $G_2$, respectively.
These, in turn, can be translated into the magnetic and electric form factors defined in the following way:
\begin{alignat*}{3}
G_\text{M}(q^2)
&\equiv
G_1(q^2)+G_2(q^2)
&&=
\kappa\left(1+\frac16\langle r_\text{M}^2\rangle q^2+\mathcal{O}(q^4)\right),\\
G_\text{E}(q^2)
&\equiv
G_1(q^2)+\frac{q^2}{(M_{\Sigma^0}+M_\Lambda)^2}\,G_2(q^2)
&&=
\frac16\langle r_\text{E}^2\rangle q^2+\mathcal{O}(q^4)\,.
\end{alignat*}
Above, we also showed the expansion of these form factors at vanishing photon virtuality $q^2$, which serves as the definition of hadronic quantities of our interest, i.e., besides $\kappa$, which is related to the magnetic moment, we have introduced $\langle r_\text{M}^2\rangle$ and $\langle r_\text{E}^2\rangle$, the magnetic and electric radii, respectively.
The effects of the electric radius are suppressed compared to its magnetic counterpart, so the matrix element squared can be in a very good approximation expressed solely in terms of the magnetic form factor:
\begin{equation}
\overline{|\mathcal{M}^\text{LO}(x,y)|^2}
\simeq
2e^4|G_\text{M}(\Delta_M^2x)|^2
\frac{(1-x)}{x}
\bigg(1+y^2+\frac{\nu^2}{x}\bigg)\,.
\label{eq:MLO2}
\end{equation}
Above, we have used kinematical variables $x$ and $y$,
\begin{equation}
x\equiv\frac{(p_{e^+}+p_{e^-})^2}{(M_{\Sigma^0}-M_\Lambda)^2}\,,\quad
y\equiv\frac{2\,p_{\Sigma^0}\cdot(p_{e^+}-p_{e^-})}{\lambda^{\frac12}(p_{\Sigma^0}^2,p_\Lambda^2,(p_{e^+}+p_{e^-})^2)}\,,
\end{equation}
where, importantly, $x$ denotes the normalized invariant mass of the electron--positron pair squared, $\lambda$ denotes the K\"all\'en's triangle function and $\Delta_M\equiv M_{\Sigma^0}-M_\Lambda$.
In Eq.~\eqref{eq:MLO2}, notice that the low-$x$ region is the dominant one.

\section{Radiative corrections}

\begin{figure}[t]
\vspace{-2mm}
\centering
\subfloat[][]{
\includegraphics[width=0.23\textwidth]{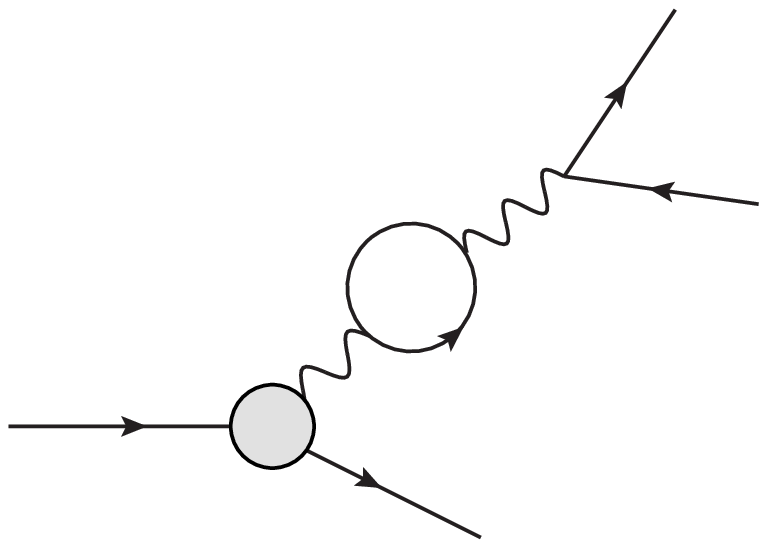}
\label{fig:virta}
}
\subfloat[][]{
\includegraphics[width=0.23\textwidth]{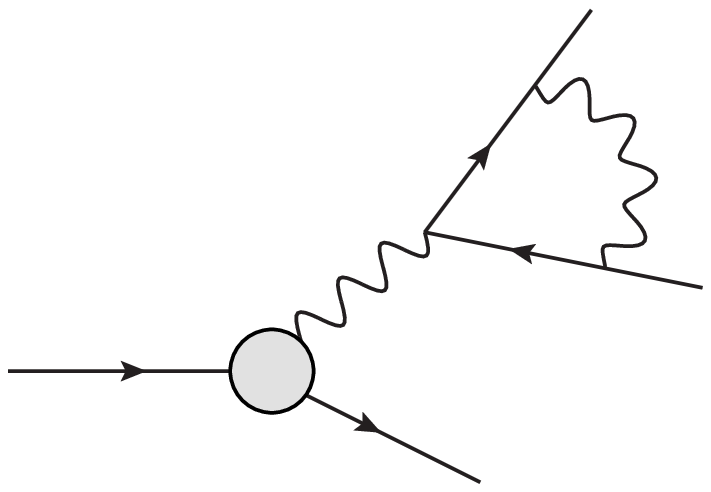}
\label{fig:virtb}
}
\subfloat[][]{
\includegraphics[width=0.23\textwidth]{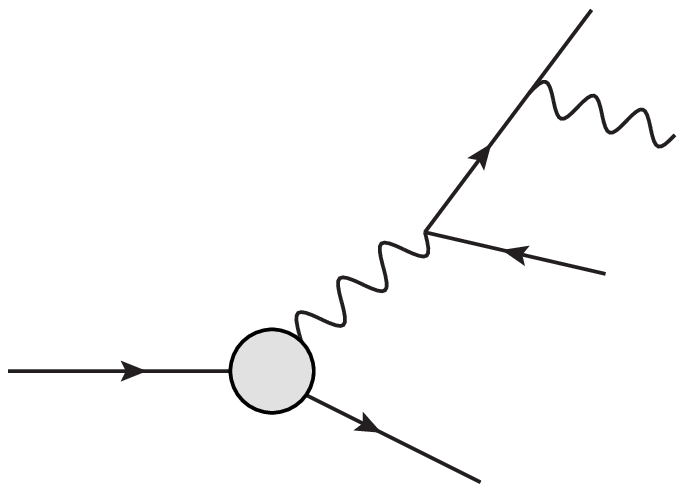}
\label{fig:BS}
}
\subfloat[][]{
\includegraphics[width=0.23\textwidth]{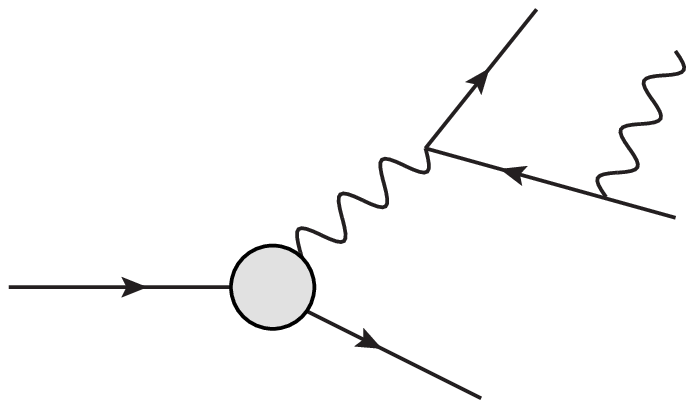}
\label{fig:BS2}
}

\subfloat[][]{
\includegraphics[width=0.23\textwidth]{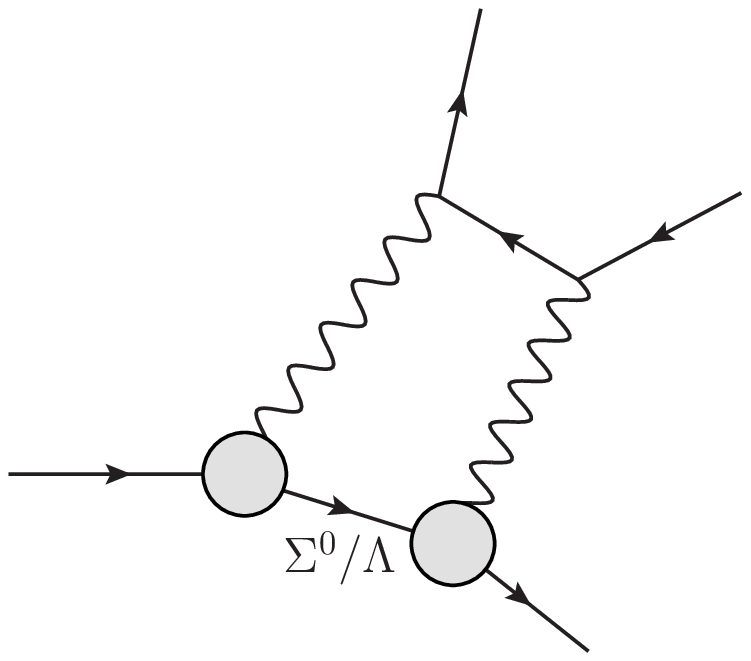}
\label{fig:1gIR}
}
\subfloat[][]{
\includegraphics[width=0.23\textwidth]{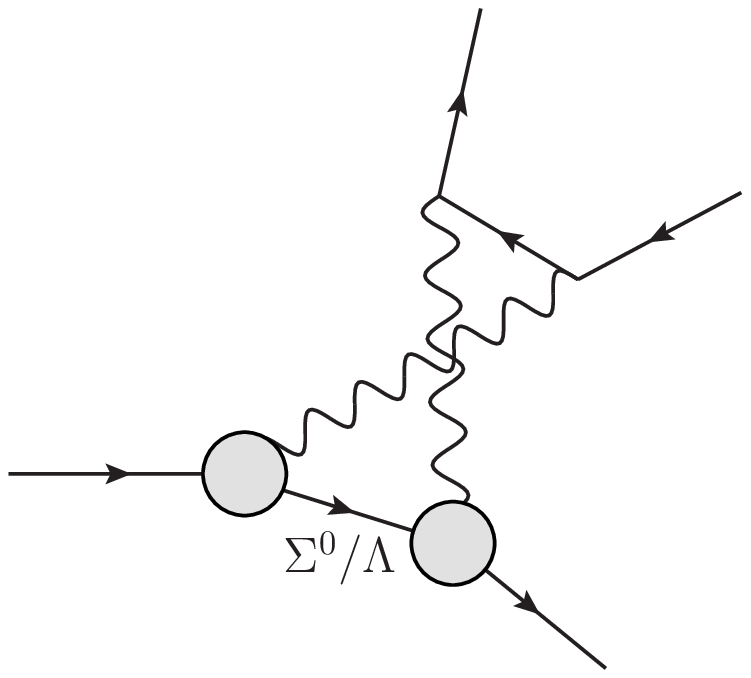}
\label{fig:1gIR2}
}
\subfloat[][]{
\includegraphics[width=0.23\textwidth]{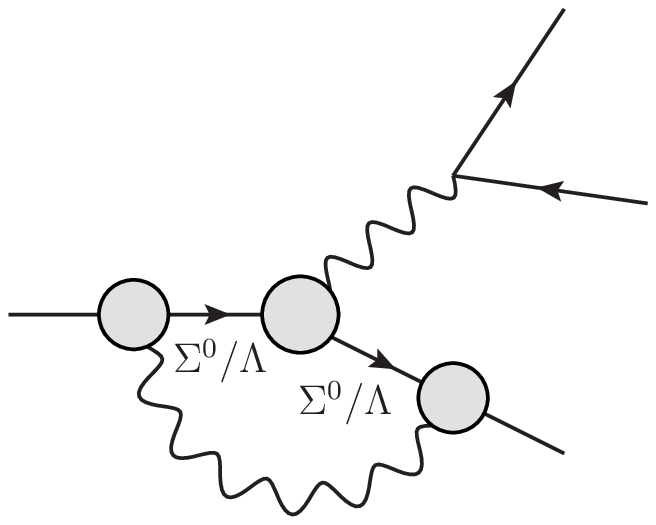}
\label{fig:virt_bar}
}
\vspace{4mm}
\caption{
\label{fig:diagrams}
NLO QED radiative corrections for the decay $\Sigma^0\to\Lambda e^+e^-$: a) lepton-loop vacuum-polarization insertion, b) correction to the QED vertex, c) \& d) bremsstrahlung, e) \& f) one-loop one-photon-irreducible (1$\gamma$IR) contributions, g) $\Sigma^0\Lambda\gamma$ vertex correction.
In the 1$\gamma$IR contribution each diagram comes in two variants: with $\Sigma^0$ or $\Lambda$ exchanged.
Similarly, there are four diagrams contributing to the transition-form-factor correction g).
}
\end{figure}

The radiative corrections to the differential decay width were already calculated in the past \cite{Sidhu:1972rx}, however, only in the soft-photon approximation.
This means that the low-$x$ region was not very well covered and the absence of the hard-photon corrections resulted in correction values which were negative all over the Dalitz plot. This was clearly in contrary to the fact that, as the authors of Ref.~\cite{Sidhu:1972rx} already knew, the overall correction to the decay rate is positive.

Such a situation was indeed unsatisfactory, so we decided to revisit the calculation and provide inclusive radiative corrections beyond the soft-photon approximation~\cite{Husek:2019wmt}.
Not only did we focus on the complete treatment of the bremsstrahlung contribution (diagrams (c) and (d) in Figure~\ref{fig:diagrams}, but we decided to explicitly calculate those contributions which were considered in the literature to be negligible --- the one-photon irreducible contribution (Figures~\ref{fig:1gIR} and \ref{fig:1gIR2}) and the correction to the $\Sigma^0\Lambda\gamma$ vertex (Figure~\ref{fig:virt_bar}) --- to verify this is the case.

As already mentioned earlier, our goal was to calculate radiative corrections for the inclusive process.
Consequently, regarding the bremsstrahlung contribution, the first (but essential) natural step to do is to perform the low-energy expansion of the form factors appearing in this contribution:
\begin{align}
G_\text{M}\big((k+q_1+q_2)^2\big)
&\simeq G_\text{M}\big((q_1+q_2)^2\big)\bigg\{1+\frac16\langle r_\text{M}^2\rangle[2k\cdot(q_1+q_2)]\bigg\}\,,
\\
G_\text{E}\big((k+q_1+q_2)^2\big)
&\simeq G_\text{E}\big((q_1+q_2)^2\big)\bigg\{1+\frac{2k\cdot(q_1+q_2)}{(q_1+q_2)^2}\bigg\}\,.
\end{align}
This allows us not only to integrate over the energy and emission angle of the bremsstrahlung photon, but the above expressions also contain form factors in the same form as they appear in the leading-order expression.
This is further useful for potential cancellations.

Moving forward to the one-loop corrections including additional hadronic form factors, simply from the loop-momenta-power counting, a proper inclusion of the form factors in the corresponding vertices is essential to regulate the potentially unwanted behaviour in the ultraviolet (UV) region.
It turns out that for the one-photon-irreducible contribution, the finite result is achieved already in the simplest case when the constant form factors are used: $G_\text{E}(q^2)=0$ and $G_\text{M}(q^2)=\kappa$, which translates into
\begin{equation}
G_1(q^2)=\kappa\,\frac{q^2}{q^2-M_V^2}\,,\quad
G_2(q^2)=-\kappa\,\frac{M_V^2}{q^2-M_V^2}\,.
\end{equation}
This is unfortunately not sufficient for the treatment of the correction to the $\Sigma^0\Lambda\gamma$.
Here, a model with stronger UV suppression needs to be taken into account.
One such example could be
\begin{equation}
G_1(q^2)=\kappa\left(3-\frac{M_V^2\langle r_\text{M}^2\rangle}{6}\right)\frac{q^2M_V^4}{(q^2-M_V^2)^3}\,,\quad
G_2(q^2)=-\kappa\,\frac{M_V^6}{(q^2-M_V^2)^3}\,.
\end{equation}
This ansatz also satisfies the high-energy behaviour dictated by the Brodsky--Lepage scaling rules~\cite{Lepage:1980fj}.
Finally, using the above-stated models, one can show via explicit calculation that the contributions under consideration are indeed negligible.
Moreover, in the case of the two-photon exchange diagram, one can show that using the above models one gets numerically compatible results.
This finding is very soothing since we don't need to be worried about any strong model-dependence of our results and conclusions.

\section{Results}

\begin{figure}[!t]
\centering
\includegraphics[width=0.75\textwidth]{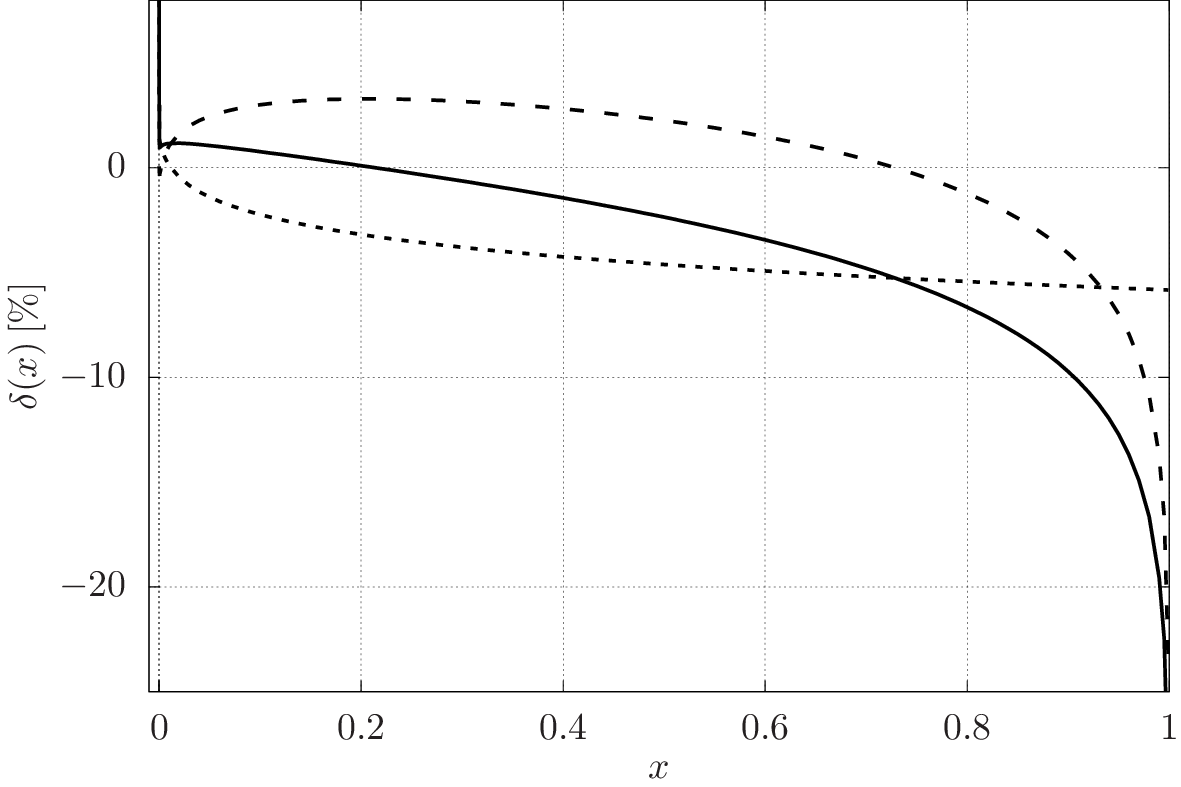}
\caption{\label{fig:deltax}
The total NLO correction for the decay $\Sigma^0\to\Lambda e^+e^-$ (solid line) in comparison to its constituents:
The virtual correction is depicted as a dotted line, 
the bremsstrahlung is shown as a dashed line.
}
\end{figure}

Studying the resulting radiative corrections for the Dalitz plot immediately reveals that compared to the soft-photon approach, there is now a region (when both the $x$ and $y$ are small) with positive values.
This is further reflected in the radiative corrections to the electron--positron-pair invariant-mass spectrum; see Figure~\ref{fig:deltax}.
Here one can see that the correction is positive for $x\lesssim0.24$.
This is sufficient to flip the sign of the overall correction, since the low-$x$ region dominates the differential decay width.

After we integrate the corrected differential decay width over the Dalitz plot, we are in a position to predict the value for the following ratio:
\begin{equation}
R
\equiv\frac{\Gamma(\Sigma^0\to\Lambda e^+e^-)}{\Gamma(\Sigma^0\to\Lambda\gamma)}
=5.541(2)\times10^{-3}\,.
\end{equation}
In the above quantity, the effects of the electric form factor can be again neglected.
Moreover, the normalization of the magnetic form factor (related to the magnetic moment) drops out in this ratio.
Regarding hadronic parameters, we are thus only left with the magnetic form-factor slope $a\equiv\frac16\langle r_\text{M}^2\rangle\Delta_M^2$, which, being numerically small, can serve as an expansion parameter.
Thus we can write
%
$
R=R_0+aR_1+\mathcal{O}(a^2)\,,
$
%
which leads to the final result
\begin{equation}
R=[5.530(3)+0.626(2)a]\times10^{-3}\,.
\end{equation}
Above, the stated uncertainty takes into account the estimated size of the higher-order corrections.
This result is consistent with the value appearing in the classical work of Sidhu and Smith~\cite{Sidhu:1972rx}:
%
$
R_\text{S\&S}
=(5.532+0.627a)\times10^{-3}\,.
$
%
Actually, when the relevant expressions are extracted from Ref.~\cite{Sidhu:1972rx} and present values for physical constants are used, one arrives at $R_\text{S\&S}^\text{new}=(5.52975+0.62640a)\times10^{-3}$, which compares very well with our result $R=(5.52974+0.62640a)\times10^{-3}$, restricting ourselves only to the corresponding set of contributions.
This serves us as a neat cross-check, since these results were obtained using different methods.

The ratio $R$ can be further translated into the branching ratios for the two dominant decay modes simply by employing the fact that all the $\Sigma^0$ branching ratios should sum up to 1:
\begin{align}
\mathcal{B}(\Sigma^0\to\Lambda\gamma)
&\simeq\frac1{1+R}
=[99.4501(3)-0.0619(2)a]\,\%\,,\\
\mathcal{B}(\Sigma^0\to\Lambda e^+e^-)
&\simeq\frac{R}{1+R}
=[0.5499(3)+0.0619(2)a]\,\%\,.
\end{align}
Using conservative $a=0.02(2)$, $\mathcal{B}(\Sigma^0\to\Lambda\gamma)=99.449(2)\,\%$ and $\mathcal{B}(\Sigma^0\to\Lambda e^+e^-)=0.551(2)\,\%$.

Finally, we can inspect our initial guess that the NLO QED radiative corrections might compete in size with the hadronic effects and estimate the size of the correction to the magnetic form-factor slope $\Delta a$.
Taking half of the slope of the total correction to the one-fold differential decay width in the low-$x$ region (although farther from the threshold), we find
\begin{equation}
\Delta a
\equiv a_\text{(+QED)}-a
\simeq\frac12\frac{\diff\delta(x)}{\diff x}\bigg|_{x=x_0\ll1}
\approx-3.5\,\%\,.
\end{equation}
Above, $a_\text{(+QED)}$ is the measured value implicitly containing the QED radiative correction, while $a$ corresponds to the purely hadronic quantity.
It thus turns out that the $\Delta a$ estimated above is in size twice as large as the estimate on the slope itself ($a\approx1.8(3)\,\%$~\cite{Kubis:2000aa}).
Translating this discussion from slopes to magnetic radii, the ``measured'' radius $\langle r_\text{M}^2\rangle_\text{(+QED)}=\langle r_\text{M}^2\rangle+\frac6{\Delta_M^2}\Delta$ is expected to be negative since $\frac6{\Delta_M^2}\Delta a\approx-35\,\text{GeV}^{-2}$, while in general for hadronic radii $\langle r^2\rangle\le(1\,\text{fm})^2\approx25\,\text{GeV}^{-2}$.

\section{Summary}

We calculated the complete set of NLO QED radiative corrections to the differential width of the $\Sigma^0$ Dalitz decay, i.e., the correction relating the QED LO calculation of the $\Sigma^0\to\Lambda e^+e^-$ process with the measurement in which, in addition, arbitrary many photons are allowed in the final state.
In particular, we calculated the lepton bremsstrahlung beyond the soft-photon approximation, the two-photon-exchange contribution and the correction to the $\Sigma^0\Lambda\gamma$ vertex.
We checked that the latter two topologies involving other hadronic form factors can be safely neglected.
We were thus able to present model-independent results in terms of single hadronic parameter~$a$.
Specifically, we showed precise and conservative predictions for the branching ratios of the two dominant $\Sigma^0$ decay modes and estimated the correction to the magnetic form-factor slope as $\Delta a\approx-3.5\,\%$.

\section{Acknowledgement}

This work has been supported in part by Grants No.\ FPA2017-84445-P and SEV-2014-0398 (AEI/ERDF, EU), by PROMETEO/2017/053 (GV), by Czech Science Foundation grant GA\v{C}R 18-17224S, and by Swedish Research Council grants contract numbers 2016-05996 and 2019-03779.


\providecommand{\href}[2]{#2}
\end{document}